\documentclass[journal]{IEEEtran}
\usepackage{epsfig,graphics,epsf,color,amsmath,balance,cite,amsfonts,multirow,stfloats,algorithm,algorithmic,mathrsfs,amssymb,color,subfigure,makecell}
\usepackage[dvipdfm,unicode,colorlinks, linkcolor=black,anchorcolor=black,citecolor=black]{hyperref}
\usepackage{upgreek}
\begin{document}

\title{Unlocking Potentials of Near-Field Propagation: ELAA-Empowered Integrated Sensing and Communication}

\author{
Zhenyao He,~Wei Xu,~Zhaohui Yang,~Hong Shen,~Ningning Fu,~Yongming Huang, Zhaoyang Zhang, \\and Xiaohu You

\thanks{
Zhenyao He, Wei Xu, Hong Shen, Ningning Fu, Yongming Huang, and Xiaohu You are with Southeast University, China (e-mail: \{hezhenyao, wxu, shhseu, 230238238, huangym, xhyu\}@seu.edu.cn).

Zhaohui Yang and Zhaoyang Zhang are with Zhejiang University, China (e-mail: \{yang\_zhaohui, ning\_ming\}@zju.edu.cn).
}}

\maketitle

\begin{abstract}
The exploration of extremely large antenna arrays (ELAAs) using high-frequency spectrum has led to a paradigm shift in electromagnetic radiation field, transitioning from the common use case of far-field propagation to near-field propagation.
This shift necessitates the modification of the conventional planar-wavefront approximation to more accurate spherical waves, exerting a profound impact on wireless transmission technologies encompassing communication and sensing.
Concurrently, integrated sensing and communication (ISAC) has gained prominence in the context of the sixth-generation (6G) wireless networks owing to its ability to cater to the ever-increasing demands of future networks.
In line with this evolving trend, this article presents a systematical investigation on ELAA-empowered near-field ISAC. We begin by introducing the fundamentals of near-field propagation with an emphasis on its double-edged effects to near-field communications. Then, we turn to near-field sensing and expound upon various typical applications. Following the separate elaborations on communications and sensing, we articulate in-depth advantages of ELAA-empowered ISAC in near field, particularly including featured opportunities arising from the dual-functional integrations, potential ISAC applications benefiting from the additional degrees-of-freedom in near field, and enablements of other complementary technologies. Finally, we outline key technical challenges that merit further exploration in the realm of ELAA-empowered near-field ISAC.
\end{abstract}

\section{Introduction}
Driven by the growing demands of the forthcoming applications in the sixth-generation (6G) wireless networks, there has been a pressing need for accurate sensing capabilities alongside communication functionalities \cite{W.XuJSTSP2022}.
In this regard, integrated sensing and communication (ISAC) has emerged as a promising solution and a prospective technique for 6G \cite{J.A.Zhang,F.Liu}, which has also been envisioned as a potential usage scenario of International Mobile Telecommunications (IMT) for 2030 and beyond \cite{ITU}. By enabling radar sensing and communication within a unified system,
ISAC introduces new avenues for enhancing spectrum efficiency with reduced hardware costs and also facilitates symbiotic advancements in both sensing and communication through their close collaboration \cite{ISAC-OTFS}.

\begin{table*}[t]
\caption{Summary of Communication, Sensing, and ISAC in the Near-field Region}
    \centering
    \setlength{\abovecaptionskip}{-4pt}
	\begin{tabular}{|c|l|l|l|}
\hline
\textbf{Functions} & \multicolumn{1}{c|}{\textbf{Featured benefits}}  & \multicolumn{1}{c|}{\textbf{Primary challenges}} & \multicolumn{1}{c|}{\textbf{Potentials with typical applications}} \\
\hline
\multirow{2}*{\textbf{Communication}} & \multirow{2}*{$\bullet$ Spatial DoF enhancement}  & \begin{tabular}[c]{@{}l@{}} $\bullet$ Distance-dependent channel acquisition \end{tabular} & \begin{tabular}[c]{@{}l@{}} $\bullet$ MIMO spatial multiplexing gain and capacity \\ enhancement \end{tabular} \\
  & & $\bullet$ Mismatch of the beam-squint effect & \begin{tabular}[c]{@{}l@{}} $\bullet$ Multiuser capacity improvement\end{tabular} \\
\hline
\multirow{5}*{\textbf{Sensing}} &\multirow{5}*{ \begin{tabular}[c]{@{}l@{}} $\bullet$ Higher resolution, accuracy, \\and sensitivity \end{tabular}}   & \multirow{5}*{$\bullet$ Complex signal processing techniques} & \begin{tabular}[c]{@{}l@{}} $\bullet$ High-accuracy target characteristic measurements \\ and visual imaging \end{tabular} \\
  & & & \begin{tabular}[c]{@{}l@{}} $\bullet$ Single-anchor-based localization and tracking  \end{tabular}\\
  & & & \begin{tabular}[c]{@{}l@{}} $\bullet$ Parameter estimations, especially the distance \\ estimation, over near-field channels \end{tabular}\\
\hline
\multirow{8}*{\textbf{ISAC}} &\multirow{4}*{ \begin{tabular}[c]{@{}l@{}} $\bullet$ Mutual benefits of sensing \\ and communication \end{tabular}}  & \multirow{4}*{ \begin{tabular}[c]{@{}l@{}} $\bullet$  Fundamental theory evolution \\ and performance analysis \end{tabular}}&  \begin{tabular}[c]{@{}l@{}} $\bullet$ Localization by reusing communication signals \end{tabular} \\
&    &   & \begin{tabular}[c]{@{}l@{}} $\bullet$  Sensing-communication mutual interference \\ mitigation in ISAC \end{tabular}\\
&\multirow{2}*{ \begin{tabular}[c]{@{}l@{}} $\bullet$ New opportunities provided \\ by the enhanced spatial DoFs \end{tabular}}    &\multirow{2}*{ \begin{tabular}[c]{@{}l@{}} $\bullet$ Hardware implementation and \\ transmission design with ELAA \end{tabular}} & \begin{tabular}[c]{@{}l@{}} $\bullet$ Physical-layer security improvement when Eve \\ and the target are in the same direction   \end{tabular}\\
&     &  & \begin{tabular}[c]{@{}l@{}} $\bullet$ Wider coverage and ISAC performance \\ enhancement via multi-station cooperation\end{tabular} \\
\hline
	\end{tabular}%
\label{table1}
\end{table*}

To achieve a substantial enhancement in data transmission rates for 6G, the effective utilization of abundant spectral resources within the high-frequency spectrum, encompassing millimeter wave (mmWave) and terahertz (THz) bands, is imperative.
Concurrently, the pursuit of higher spectral efficiency represents another key aspect, driving the emergence of various cutting-edge technologies.
The extremely large antenna arrays (ELAAs) emerge as a crucial contender in this regard.
ELAAs not only serve to compensate for the considerable path loss experienced by high-frequency signals but also yield significant spatial multiplexing gains. In practical networks, ELAAs can be deployed through flexible manners, such as centralized ultra-massive multiple-input multiple-output (UM-MIMO), distributed cell-free architecture, multi-station cooperation, and reconfigurable intelligent surface (RIS).

Utilizing ELAAs at high-frequency spectrum instigates a fundamental metamorphosis in electromagnetic (EM) characteristics. That is, the EM radiation field transitions from the far-field region to the near-field region, necessitating a more precise representation of the signal wavefront as spherical propagation rather than a simplistic planar approximation. Concretely, Rayleigh distance serves as the commonly employed criterion to demarcate the boundary between the near-field and far-field propagations, which is proportional to the product of the squared array aperture and the carrier frequency. The substantial increase in both array aperture and frequency expands the Rayleigh distance, rendering it no longer negligible as observed in previous communication networks. As a crucial consequence, the near-field spherical-wavefront propagation introduces the domain of distance, in addition to the predominantly considered angle domain in far-field scenarios, which further opens up the possibility to focus signals at specific locations and engenders a profound impact on wireless transmission technologies.

Recent studies have shed light on the multifaceted implications of near-field propagation in ELAA-aided wireless communications, including a spectrum of both opportunities and challenges \cite{NFComMag}. In the realm of radar sensing, where the near field represents a more frequently visited subject, the adoption of spherical-wave propagation and the abundant spatial multiplexing gain in near-field regions have exhibited enhancements in sensing resolution and accuracy \cite{imaging1,tracking}.
While communication and sensing in near-field regions have already been well studied, research on near-field ISAC has only recently begun.
Therefore, this article aims to comprehensively explore the behaviors of ELAA-empowered ISAC within the near-field region and elucidate the distinguishing characteristics it presents.
Specifically, we first provide a concise overview of the fundamental principles underlying near-field propagation and its dual impacts on wireless communications. Subsequently, we elaborate on several application scenarios related to near-field sensing.  Thereafter, we explore the novel opportunities that ELAA-empowered near-field ISAC can offer, followed by the potential ISAC applications enhanced by the additional degrees-of-freedom (DoFs), and the augmentation of other enabling technologies. Finally, we present an in-depth discussion of key open research directions and challenges.

\section{Near-Field Propagation and Communications}
Near-field communications have attracted considerable attentions recently.
In this section, we briefly review the fundamentals of near-field propagation utilizing high-frequency ELAAs and highlight the distinctive potentials and challenges to wireless communications.

\subsection{Fundamentals of Near-Field Propagation}
In near-field regions, the introduction of a spherical wavefront brings forth a crucial characteristic that encompasses both the conventional angular information concentrated in far-field scenarios and an additional \textit{dimension of distance}.
Furthermore, in comparison to far-field beamforming which enables the transmission signal to be directed towards a specific angular direction across varying distances, commonly known as \textit{beamsteering}, near-field beamforming facilitates the concentration of the signal at a precise location within the polar domain, which is known as \textit{beamfocusing}.

\begin{figure*}[t]
\centering
\setlength{\abovecaptionskip}{-2pt}
      \epsfxsize=7.0in\includegraphics[scale=0.46]{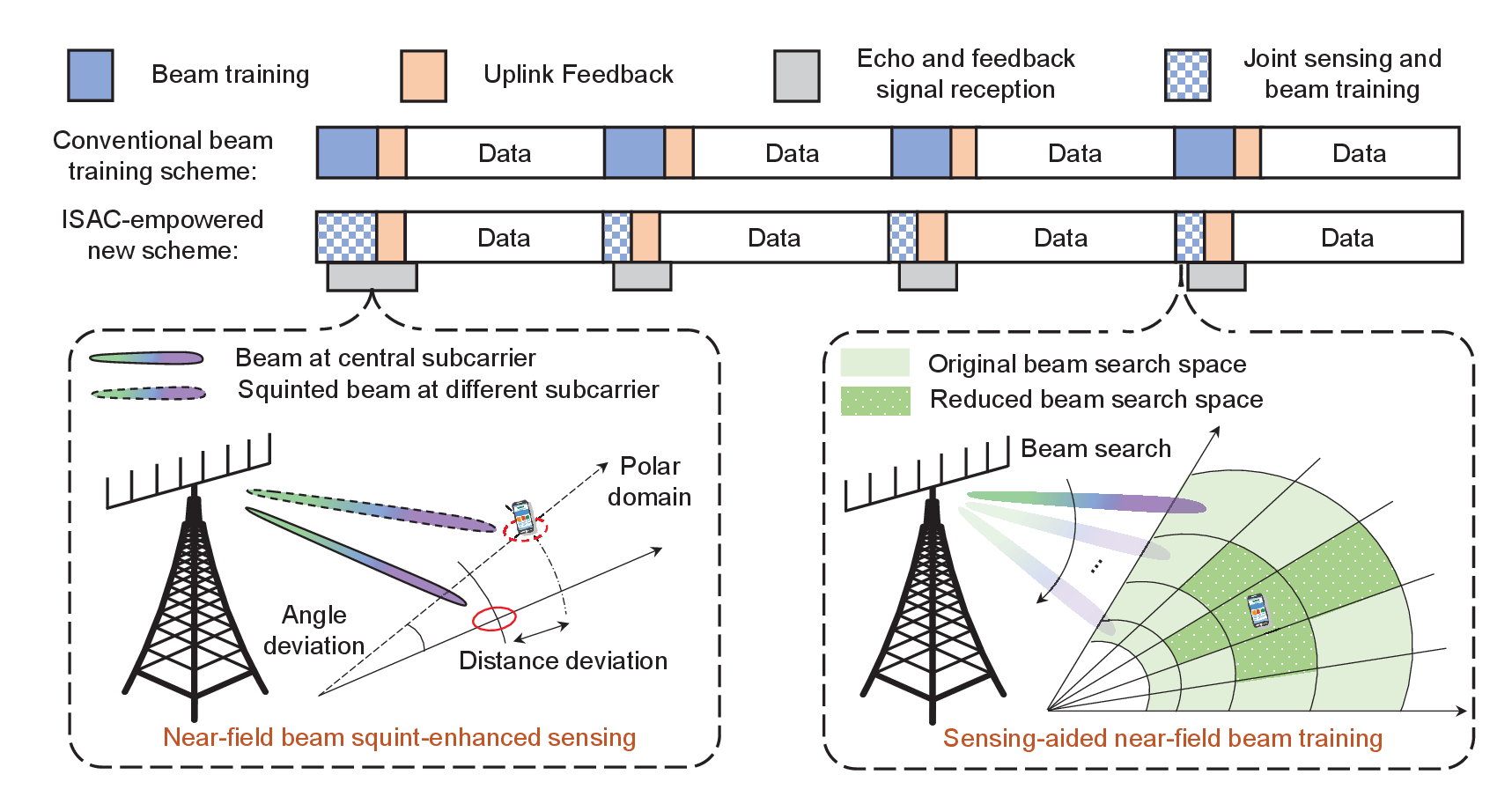}
      \caption{ISAC-empowered near-field transmission scheme: near-field beam squint-enhanced sensing and sensing-aided near-field beam training.}\label{fig:oppor}
\end{figure*}

\subsection{Wireless Communications in Near-Field Regions}
Near-field communications, like in \cite{NFComMag}, exhibit substantial exploration potentials while also presenting unique challenges.

\subsubsection{Potentials}
The primary advantage stems from the distance dimension in near-field propagation, which substantially augments spatial DoFs and enhances the performance of wireless communications.
For instance, for a point-to-point MIMO channel relying on single line-of-sight propagation, the rank-deficiency issue encountered in far-field scenarios is mitigated in near-field communications due to the additional distance-dependent DoFs. In multiuser scenarios, far-field communications cannot distinguish users located at the same angle resolvable by the MIMO array. Near-field communications overcome this limitation by concentrating beams on specific grid points rather than solely angles, enabling further enhancements in the capacity of high-frequency multiuser MIMO.

\subsubsection{Challenges}
Two demanding challenges arise in near-field communications: channel acquisition and beam squint.
While high-frequency far-field channels exhibit angle-domain sparsity, the incorporation of distance-dependent information renders existing codebooks designed for the angle domain in far-field scenarios inadequate for the near field.
On the other hand, ELAAs commonly utilize hybrid array architectures to balance implementation costs.
However, subcarrier-independent analog beamforming fails to align with frequency-dependent spatial steering vectors across various subcarriers, yielding a phenomenon known as beam squint. The near-field beam squint focuses beamforming points on different angle-distance directions at different frequencies, severely degrading communication performance.

\section{Wireless Sensing in Near-Field Regions}
The massive diversity and refined spatial resolution achieved by ELAA, as well as the ultra-wide bandwidth, offer significant advantages for radar sensing.
In contrast to communications, the near-field concept is more familiar for sensing and has been investigated over the past few decades. In this section, we elaborate on several typical use cases of near-field sensing.
\subsection{Near-Field Radio Frequency Imaging}
Active radio frequency (RF) imaging technology finds applications in a multitude of scenarios, including environmental monitoring, medical diagnosis, and security detection, etc.
Generally, RF imaging can be utilized to understand the properties and reflection characteristics of a target, such as radar cross-section measurement and material characterization, as well as to generate recognizable images of objects that may be visually obstructed, as exemplified by the emerging THz imaging techniques.
By harnessing the capabilities of near-field signals, RF imaging offers numerous advantages, e.g., improved resolution, high sensitivity, and near-real-time image acquisition \cite{imaging1}.
For instance, in contrast to conventional far-field visual imaging, which faces spatial resolution limitations typically on the order of the wavelength, near-field imaging enables the detection of sub-wavelength scale details and offers the potential for significantly higher spatial resolution.

\subsection{Near-Field Localization based on Spherical Wavefront}
Localization, as well as tracking, has emerged as a pivotal application in wireless networks. In general, device localization can be accomplished through multi-anchor collaboration and single-anchor computation \cite{F.Liu}. In a multi-anchor scenario, the distance or the time of arrival of the target needs to be estimated and then gathered from multiple nodes to compute the target location. Single anchor-based localization requires a joint estimation of time and angle of arrivals. In such schemes, time-related information heavily relies on precise synchronization among the involved nodes, while distance estimation can be compromised in cluttered environments.
In near-field regions, leveraging the combined potential of high frequencies and ELAA bestows significant enhancements to localization and tracking capabilities. Precisely, the spherical-wave characteristic of near-field propagation allows for the acquisition of additional target information. One example is to extract the target position directly from the curvature of arrival (CoA) information of the spherical wavefront \cite{tracking}.

\subsection{Parameter Estimations of Near-Field Channels}
Estimating channel parameters, such as angle and Doppler frequency of a target, constitutes another essential aspect of wireless sensing. When considering spherical-wavefront characteristic in near-field channels, a notable distinction arises. That is, the distance information is involved and can be directly extracted from received echoes, which is typically unavailable in far-field cases.
Specifically, the Cram\'er-Rao bound (CRB) for distance estimation, conventionally infinite in far-field planar-wave propagation, becomes finite in near-field regions \cite{CRBsensing}, confirming the capability for precise distance discrimination in near-field sensing facilitated by ELAA.

\begin{figure*}[t]
\centering
\setlength{\abovecaptionskip}{-2pt}
      \epsfxsize=7.0in\includegraphics[scale=0.46]{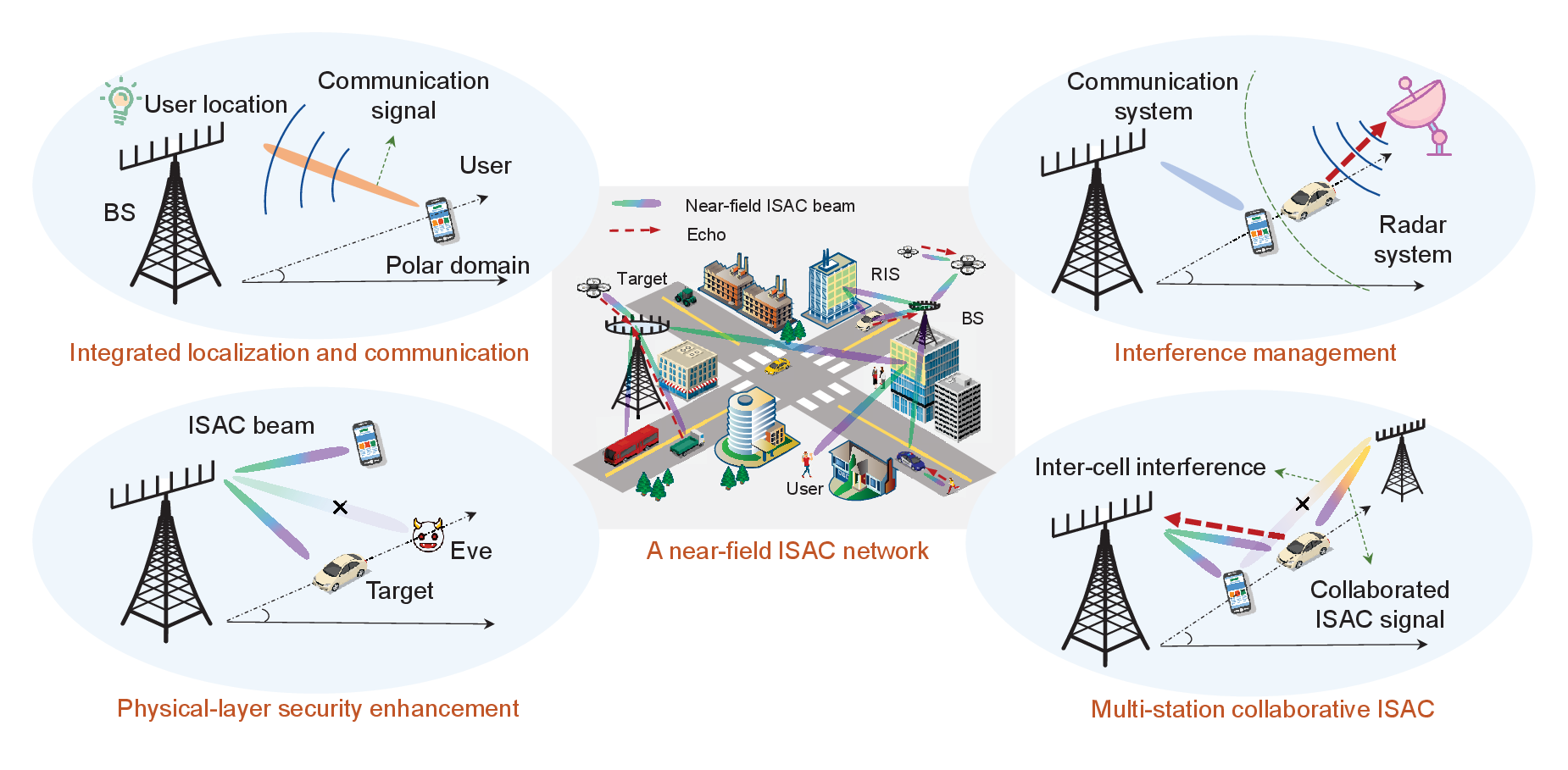}
      \caption{An illustration of near-field ISAC network and four representative potential applications: integrated localization and communication, interference management, physical-layer security enhancement, and near-field multi-station collaborative ISAC.}\label{fig:usecase}
\end{figure*}

\section{ELAA-Empowered Near-Field ISAC}
Following the separate exposition on the distinctive features of near-field communications and sensing, with a summary and comparison provided in Table~\ref{table1}, this section focuses on exploring the potentials of ELAA-empowered near-field ISAC.

\subsection{Solution and New Viewpoint for Near-Field ISAC}
ISAC provides promising solutions to address the challenges encountered in near-field communications. Specifically, through the collaborative utilization of sensing, the difficulty associated with channel estimation can be alleviated. Additionally, beam squint exhibits distinct behaviors under the ISAC framework, offering new opportunities. To illustrate these features more explicitly, we present an example of the ISAC-empowered new transmission scheme in Fig.~\ref{fig:oppor}, accompanied by the descriptions as follows.

\subsubsection{Sensing-Aided Near-Field Beam Training}
The utilization of a polar-domain codebook-based beam training technique has proven to be effective for near-field channel estimation \cite{NFComMag}. Nevertheless, conducting an exhaustive search throughout the entire codebook for beam measurements generally leads to unbearable latency, which is further exacerbated in ELAA-based near-field scenarios where the codebook encompasses both angle and distance domains and contains a larger number of codewords. Fortunately, the integration of sensing within ISAC offers a substantial alleviation of this issue.
Specifically, leveraging the results obtained from pre-conducted sensing operations, such as user angle and distance estimations, prior information pertaining to the user channel can be acquired.
Consequently, as depicted in the bottom right of Fig.~\ref{fig:oppor}, the need for beam training can be either obviated or, at the very least, the search space for candidate beams can be significantly reduced.
In addition, novel and advanced techniques \cite{ISAC-OTFS,tracking} enable the prediction of channel variations, facilitating the communication with mobile users.

\subsubsection{Near-Field Beam Squint-Enhanced Sensing}
The near-field beam-squint effect introduces challenges to communications due to mismatch at the user. However, it can be advantageous for the sensing purpose from the following two perspectives.
Firstly, as illustrated in the bottom left of Fig.~\ref{fig:oppor}, with the squinted beams, the users positioned within the service range measure the received array gain of different subcarriers, and feed back the subcarrier index associated with maximum gain to the transmitter.
Then, by exploiting the mathematical relationship between subcarrier frequency and its corresponding squinted locations in near-field regions, the transmitter can utilize this feedback to determine user locations \cite{Beam-Split}. Moreover, this approach allows for the simultaneous sensing of multiple targets situated at different locations, reducing the beam scanning overhead.
Secondly, by employing the same analog beamforming vector, wider coverage is achieved as beams are focused towards different angle-distance locations across various subcarriers, which is particularly advantageous for extended target sensing.

Furthermore, the aforementioned two new solutions can be integrated within an ISAC-empowered new transmission scheme in a collaborative manner, as exemplified in Fig.~\ref{fig:oppor}. Specifically, at the initiation of a transmission stage, the base station (BS) conducts wide-coverage joint sensing and beam training. Utilizing the uplink feedback from the user, the BS not only determines the most suitable transmit beam aligned with the user but also captures additional user information such as angle and distance, leveraging the benefits of beam squint-enhanced sensing.
Subsequent blocks utilize these prior sensing information related to user location and channel, enabling the BS to streamline beam training overhead and reduce latency. This reflects the benefits of sensing-aided near-field beam training.

\subsection{Benefits and Potential Applications of Near-field ISAC}
For ISAC, the additional spatial DoFs of the near-field signals, including the introduction of distance dimension and the beamfocusing capability, and the high spatial resolution facilitated by ELAA offer novel opportunities.
Fig.~\ref{fig:usecase} exemplifies four representative potential applications, each accompanied by a description outlined below.

\subsubsection{Integrated Localization and Communication}
In future wireless networks, the capability of access points or BSs to support localization and tracking services alongside communication, known as integrated localization and communication, is highly anticipated.
Leveraging the near-field spherical-wave propagation, localization can be integrated into communication networks equipped with ELAAs in an efficient manner.
For instance, as exemplified in the top left of Fig.~\ref{fig:usecase}, the user location can be directly estimated at one single station by analyzing the CoA information of the uplink signal \cite{tracking}, without requiring major modifications to existing communication infrastructure.
Furthermore, in downlink scenarios where the transmit signal serves a dual purpose of simultaneous communication and target sensing, the location of the near-field target can be determined by extracting the distance and angle information from the echoes \cite{NF-ISAC}.

\subsubsection{Interference Management}
Due to the reuse of spectral resource, mutual interference between communication and sensing is a severe issue in ISAC. One specific instance of ISAC, known as radar-communication coexistence (RCC), can be viewed as a scenario with a lower level of integration \cite{J.A.Zhang}. In RCC, the radar and communication systems share the same spectrum, yielding co-channel interference. Additionally, the incorporation of dedicated radar waveforms, devoid of any data symbols, in the ISAC transmitter for higher DoFs introduces interference at the communication receiver \cite{F.Liu}.
These instances highlight the necessity of an effective interference management approach in ISAC. As illustrated in the top right of Fig.~\ref{fig:usecase}, the beamfocusing capability and the abundant spatial DoFs inherent in near-field signals hold promise to mitigate such interference and enhance the system performance.

\subsubsection{Physical-Layer Security Enhancement}
Physical-layer security is a crucial concern in ISAC, as the transmit signal for environmental sensing exposes the embedded communication data to potential eavesdropping. For the security considerations, it is essential to ensure sufficient power emitted towards the target while minimizing leakage to prevent eavesdropping in ISAC.
It has been proven challenging to achieve this goal in far-field scenarios when the eavesdropper (Eve) is in close proximity to the target, particularly if they are located in the same direction. In this context, near-field beamfocusing offers a viable solution by focusing the signals on the target at a specific range while reducing information leakage to Eve at a different range, as demonstrated in the bottom left of Fig.~\ref{fig:usecase}.
In addition, the high spatial DoFs and resolution of near-field signals also present significant advancements for commonly employed secure beamforming and jamming techniques.

\subsubsection{Multi-Station Collaborative ISAC}
Multi-station collaborative ISAC offers wider coverage and performance enhancement for both communication and sensing through joint transmission and combination at distributed nodes.
In addition, the spatial DoFs enhanced by near-field propagation are further augmented as the number of collaborative stations increases, significantly improving the number of users and targets that can be simultaneously served and yielding higher spatial multiplexing gain and capacity.
A potential use case for collaborative ISAC is depicted in the bottom right of Fig.~\ref{fig:usecase}, where communication user receives data and suffers interference from the desired station and adjacent stations, respectively.
While the probing power directed towards the target is intended to be maximized to ensure the multistatic sensing performance.
The beamforming capability and increased DoFs offered by near-field signals prove advantageous in this regard, enabling the simultaneous suppression of interference and enhancement of probing signals towards different points.

\subsection{Enabling Techniques for Near-Field ISAC}
We present two representative enabling techniques beneficial for near-field ISAC: RIS-enabled near-field region construction and full-duplex (FD)-enhanced near-field ISAC.
\subsubsection{Active Near-Field Region Construction by RIS}
\begin{figure}[t]
\centering
\setlength{\abovecaptionskip}{-2pt}
      \epsfxsize=7.0in\includegraphics[scale=0.42]{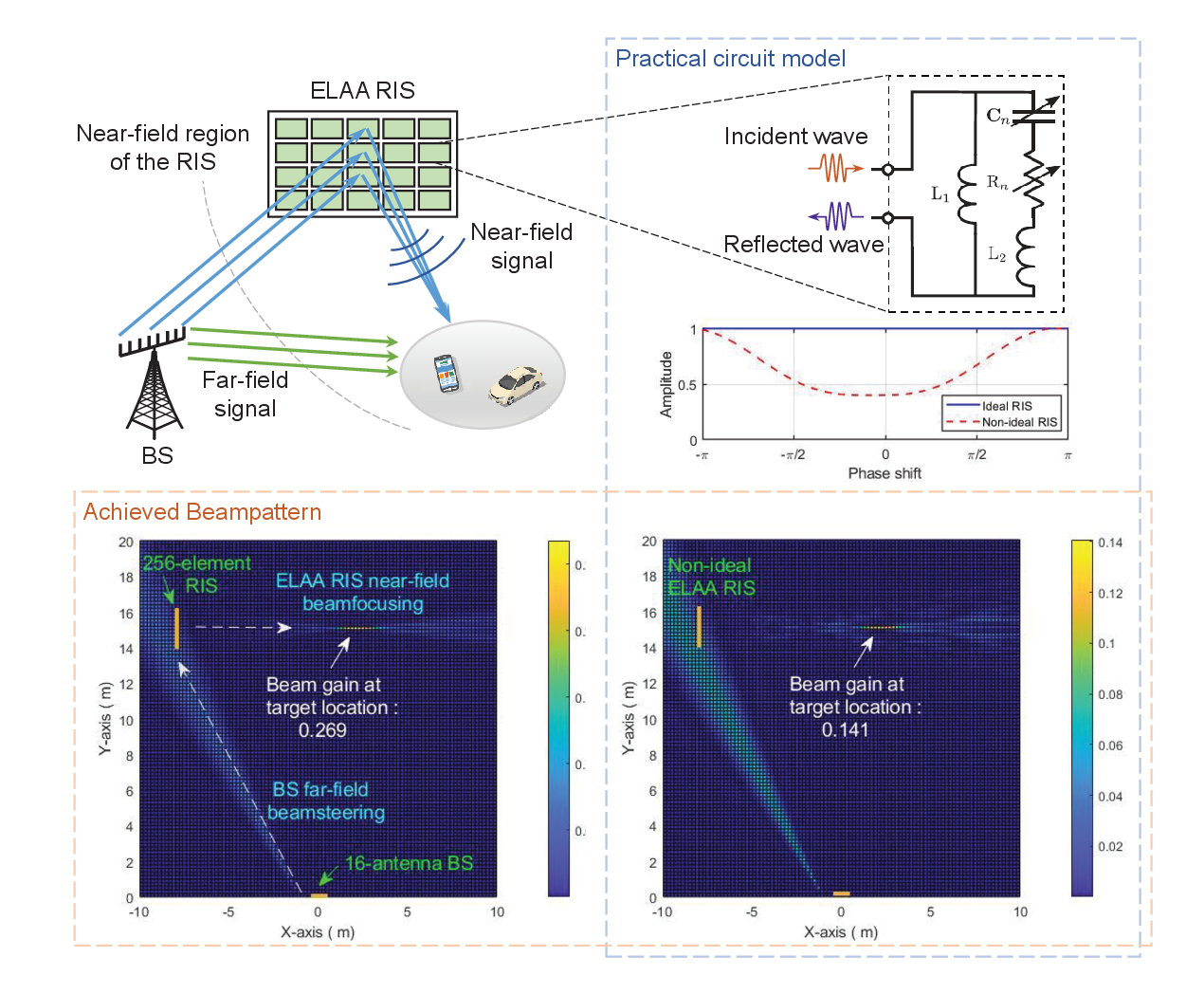}
      \caption{RIS-enabled near-field ISAC: active near-field region construction via deploying ELAA RIS and the impact of non-ideal reflection elements in RIS.
}\label{fig:RIS}
\end{figure}

The exceptional benefits of RIS, as a typical application of ELAA, have been extensively demonstrated in the domains of communication, sensing, and ISAC, primarily attributed to its ability to reconstruct wireless environments. Additionally, as exemplified in the top left of Fig.~\ref{fig:RIS}, the introduction of ELAA RIS can significantly extend the near-field range of ISAC systems, thereby encompassing users and targets within the near-field region of ELAA to leverage the advantages offered by near-field signals \cite{RIS}.
However, the commonly used ideal RIS phase shift model, which assumes full signal reflection at each element, is challenging to implement in practice, particularly for large-scale ELAA RISs. Therefore, it is necessary to explore practical circuit-based non-ideal RIS phase shift models that incorporate phase-dependent amplitude variation \cite{RISPracticalModel}, as depicted in the top right of Fig.~\ref{fig:RIS}. In such cases, employing conventional reflection coefficient design schemes that assume full reflection cannot accommodate this non-ideality and leads to noticeable performance losses, as exemplified in the bottom right of Fig.~\ref{fig:RIS}. To address this issue, advanced beamforming optimization approaches that directly consider the non-ideal RIS reflection model can be utilized.

\subsubsection{FD-Enhanced Near-Field ISAC}

In the context of near-field ISAC, FD operation warrants particular attention over its far-field counterpart. This is due to the negligible time difference between signal transmission and echo reception at a mono-static radar transceiver in near-field scenarios with limited operational range.
Consequently, the radar component necessitates operation in FD mode in near-field ISAC. In such scenarios, the self-interference (SI), referring to the signal leakage between the transceiver, emerges as a critical concern. To tackle this issue, SI cancellation techniques can be applied, encompassing natural isolation, analog cancellation, and digital cancellation, achieving an impressive total SI suppression of over 100~dB \cite{F.Liu}. With this remarkable SI cancellation capability, it is natural to extend FD operation to the communication component in near-field ISAC, enabling concurrent uplink data reception alongside downlink ISAC transmission to attain higher spectral efficiency. Notably, SI exhibits slight difference in its impact on sensing and communication. From a radar sensing perspective, preserving the echoes reflected from the targets while suppressing the direct coupling between the transceiver is essential. Conversely, in communications, both components of target reflection and direct transceiver coupling introduce interference to the data decoding process. Leveraging careful transceiver design within this FD ISAC framework \cite{FDJSAC}, substantial performance enhancements can be reached for both sensing and communication.

\subsection{Case Study}
\begin{figure}[t]
\centering
\subfigure[]{
\begin{minipage}[t]{0.9\linewidth}
\centering
\includegraphics[scale=0.5]{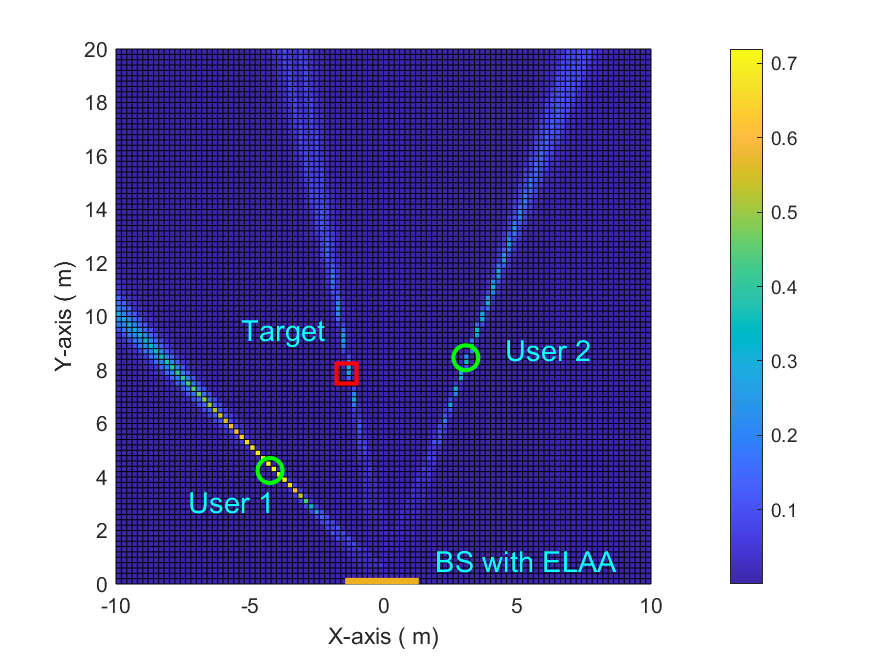}
\end{minipage}%
}
\centering
\subfigure[]{
\begin{minipage}[t]{0.9\linewidth}
\centering
\includegraphics[scale=0.5]{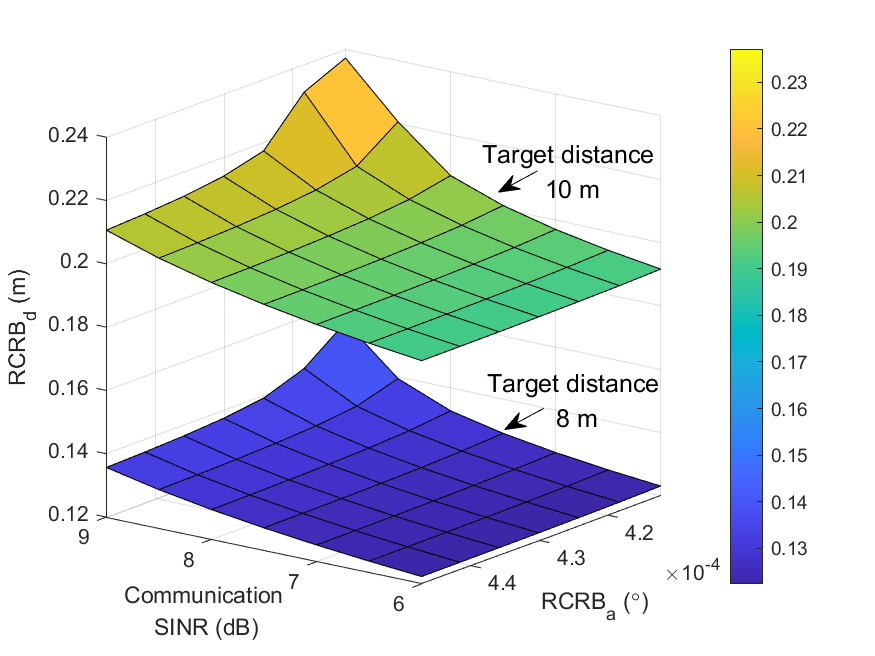}
\end{minipage}%
}%
\centering
\caption{Computer experimental examples of an ELAA-empowered near-field ISAC system: a) beampattern of near-field ISAC; b) performance of near-field ISAC, where $\text{RCRB}_{\rm a}$ and $\text{RCRB}_{\rm d}$ represent the RCRBs for angle and distance estimations, respectively.}\label{fig:casestudy}
\end{figure}

To demonstrate the potentials and advantages of ELAA-empowered near-field ISAC more clearly, we present a case study. Assume that an ISAC BS, equipped with 8 RF chains and a 128-element uniform linear array with half-wavelength spacing, communicates with two single-antenna users and simultaneously performs joint distance-angle estimation on a point target.
The locations of the two users and the target are set as $(6~\rm{m},45^\circ)$, $(9~\rm{m},110^\circ)$, and $(8~\rm{m},80^\circ)$, respectively. The carrier frequency is set at 15 GHz. The target and users fall within the near-field region of the BS.
The transmit beamforming is optimized to facilitate target sensing by minimizing the CRB on joint distance
and angle estimation, under the constraints of power limit, normalized to 10~dB, and the minimum communication signal-to-interference-plus-noise ratio (SINR) requirement, set to 5~dB.
Fig.~\ref{fig:casestudy}(a) illustrates the transmit beampattern of the BS, where three main beams are directed towards the specific locations of the users and target, indicating near-field beamfocusing.
We further evaluate and visualize the achievable sensing and communication performances in Fig.~\ref{fig:casestudy}(b). We observe that, in addition to the communication requirement, there also exists a trade-off between angle and distance estimations in near-field ISAC. That is, with a fixed requirement for communication, the root of CRB (RCRB) for distance estimation increases as the RCRB for angle estimation decreases. Furthermore, we note that an increase in target distance leads to a degradation in distance estimation accuracy. This is because, as the transmission distance increases, the wireless channel gradually transitions from the near field to the far field, therefore resulting in reduced capability for distance discrimination.

\section{Open Challenges and Future Directions}
To facilitate the practical applications, in this section we discuss several key research issues and technical challenges for ELAA-empowered near-field ISAC.

\subsection{Fundamental Theories and Performance Analysis}
\textit{Evolution of Fundamental Theories:} Certain adaptations to relevant fundamental theories are necessary in the context of near-field ISAC. For instance, the Rayleigh distance, which measures the near-field range, is defined relying on the phase discrepancy between the far-field planar approximation and the near-field spherical wave. This phase-based definition may have varying implications for sensing applications, given the round-trip nature of the radar channel. A more accurate quantification of the near-field range in ISAC by thoroughly investigating distinct characteristics of the communication and radar channels is needed.
In addition, a hybrid-field configuration, encompassing both far-field and near-field signal components, often arises in ISAC. In particular, in sensing setups employing radar transceivers with distinct array captures, the targets of interest may locate in the near-field region of one array but reside in the far-field region of the other. Advanced theories and approaches that account for the hybrid-field effect in ELAA-empowered ISAC are imperative.

\textit{Performance Trade-Offs and Limits:} The novel interplay between communication and sensing in the context of near-field ISAC necessitates a re-examination of their trade-offs and performance limits. For instance, in Fig.~\ref{fig:casestudy}(b), the CRBs associated with angle and distance estimations for a target in near-field ISAC exhibit a competitive relationship, leading to a more intricate communication-sensing trade-off when further incorporating communication metrics. As a consequent, there is a critical need to explore fundamental connections and mutual impacts between communication and sensing metrics within the near-field ISAC scenario. Furthermore, understanding these interactions is helpful to uncover their performance limits, such as constructing the Pareto-optimal boundary for near-field ISAC.

\subsection{Hardware Implementations and Transmission Designs}

\textit{Array Architectures:}
To facilitate the utilization of ELAA for near-field ISAC, novel developments in array architecture are essential.
Notably, in addressing the challenge posed by the dramatically large number of antennas in ELAA, the concept of a continuous-aperture array has emerged as a relatively novel approach. This technique allows for the denser placement of antenna elements without the constraint of half-wavelength spacing \cite{HoloMIMO,RIS}. Such an advancement results in higher spatial resolution and enhanced operability of EM waves, offering substantial benefits for near-field ISAC applications.

\textit{Hardware Implementations:}
Hardware implementation for the integration of sensing and communication in ELAA systems necessitates meticulous attention.
In general, communication and sensing, driven by distinct functional objectives, put forward different demands on their hardware components.
For example, the near-field beam-squint effect presents a dual-edged impact for communication and sensing tasks.
Leveraging this effect can enhance sensing capabilities, yet it poses challenges for communication quality and necessitates appropriate compensation strategies, such as true-time-delay (TTD)-based beamforming, to counteract its adverse impacts.
Addressing the issue of fulfilling the differing requirements for sensing and communication tasks within a same hardware platform, as well as handling the resultant hardware non-idealities and imperfections, warrants further exploration.

\textit{Transmission Designs:} For the purpose of serving both communication and sensing simultaneously, the transmission design, encompassing waveform and beamforming techniques, employed in time, frequency, code, and spatial domains has become a focal point in ISAC research \cite{J.A.Zhang,F.Liu}. With the changes of array architecture, hardware equipments, and channel characteristic, along with the significantly larger number of antennas in ELAA, the transition from far-field ISAC to near-field ISAC presents new challenges to existing transmission design techniques.

\section{Conclusion}
In this article, we investigate the potentials of ELAA-empowered near-field ISAC.
Starting with an elucidation of the principles of near-field propagation, we then present a concise overview of near-field communications and describe diverse applications of near-field sensing.
Thereafter, we envision the immense prospects of near-field ISAC facilitated by ELAA, encompassing novel approaches and new viewpoints to tackle the challenges encountered in near-field communications, exemplary applications augmented by the additional spatial DoFs, and several enabling techniques.
Additionally, we present a case study to illustrate the performance advantages of utilizing ELAA to ISAC. Finally, we shed light on the open challenges and future directions for the advancement of ELAA-empowered near-field ISAC.

\end{document}